\documentclass[fleqn,usenatbib]{mnras}

\usepackage{color,soul}

\usepackage[T1]{fontenc}
\usepackage{ae,aecompl}

\usepackage{graphicx}	
\usepackage{amsmath}	
\usepackage{amssymb}	
\usepackage[dvipsnames]{xcolor}

\usepackage{newtxtext,newtxmath}

\title[Local lensing dipole]{The effects of lensing by local structures on the dipole of radio source counts}

\author[C. Murray et al.]{
Calum Murray,$^{1}$\thanks{E-mail: calumhrmurray@gmail.com}
\\
$^{1}$Université Grenoble Alpes, CNRS, LPSC-IN2P3, 38000 Grenoble, France
}

\date{Accepted XXX. Received YYY; in original form ZZZ}

\pubyear{2021}

\begin{document}
\label{firstpage}
\pagerange{\pageref{firstpage}--\pageref{lastpage}}
\maketitle

\begin{abstract}
 Our peculiar motion in a homogeneous and isotropic universe imprints a dipole in the cosmic microwave background (CMB) temperature field and similarly imprints a dipole in the distribution of extragalactic radio sources on the sky. Each of these effects have been measured, however each of these measurements give different results for the velocity of our motion through the Universe: the radio dipole measurements finds the speed of our motion to be around three times larger than that of the CMB. Here we show the effects of the previously unconstrained lensing dipole, whereby necessarily local structures (required for large angular lensing scales) will distort the distribution of radio sources on the sky. We find that the inclusion of these effects does not reduce the tension between the CMB and radio source dipole measurements however without their inclusion future extragalactic number counts could lead to incorrect inferences of our peculiar motion. In addition we can constrain the size of the lensing dipole to be $\kappa < 3 \cdot 10^{-2}$ at the $2 \sigma$ level.
\end{abstract}




\section{Introduction}

The isotropy of the universe is an essential part of the cosmological principle; it is important that such an assumption is underpinned via careful observation. The measurement of our velocity relative to two different cosmological frames, the cosmic microwave background (CMB) and distant radio sources, find different values for the speed but are consistent for the direction of the velocity. Future extragalactic observations (SKA \citep{bacon2020cosmology}, Rubin \citep{ivezic2019lsst}, Euclid \citep{laureijs2011euclid}) should answer the question as to whether such a discrepancy is due to physical phenomena or a systematic observational issue. 

Physical implications of a discrepancy between our relative motion to the CMB and our relative motion towards radio source counts are complex. In general it would bring into question the assumption of a homogeneous universe. However given the many successes of the $\Lambda$CDM model it is interesting to consider other possibilities for the discrepancy. For example super-Hubble isocurvature modes could contribute significantly to the intrinsic CMB dipole \citep{langlois1995dipole}. This super-Hubble mode contribution would have to be $10^2$ larger than the $\Lambda$CDM intrinsic dipole, it is unclear if such a large contribution is feasible.

With the assumption that the dipole anisotropy observed in the CMB is purely kinematic, a reasonable assumption as an intrinsic dipole should have an amplitude on the order of $\Delta T/T \sim 10^{-5}$ opposed to the measured amplitude of $\Delta T/T \sim 10^{-3}$, the CMB place by far the strongest constraints on the kinematic dipole and our motion relative to the CMB rest frame finding the speed to be $369.82 \pm 0.11$
[km/s] in the direction $l= 264.021^{\circ} \pm 0.011^{\circ}$, $b=48.253^{\circ} \pm 0.005^{\circ}$ \citep{collaboration2020planck}. This constraint is made by measuring the Doppler shift of the CMB monopole due to the observer motion.

The kinematic dipole may also be constrained through the aberration and modulation of the primordial CMB temperature fluctuations (\cite{challinor2002peculiar}, \cite{burles2006detecting}). Also using Planck weaker constraints have placed the CMB kinematic dipole using the primordial anisotropies with $384 \pm 78$ (stat.) $\pm 114$ (syst.) [km/s] \citep{aghanim2014planck}.

Distant radio sources are affected by our relative motion towards in a similar manner \citep{ellis1984expected}. The observer motion induces a dipolar pattern into the assumed isotropic distribution of radio sources on the sky via Doppler boosting which induces a frequency shift to incoming photons and a modulation of their intensity and aberration in which incoming photons are deflected towards the direction of our motion. Many constraints have been made on the kinematic dipole using this method and each of them find a consistent direction with the CMB dipole but a larger amplitude  (\cite{blake2002detection},\cite{crawford2009detecting}, \cite{singal2011large}, \cite{gibelyou2012dipoles}, \cite{rubart2013cosmic},\cite{tiwari2015dipole},\cite{colin2017high}  ). In more detail, \cite{tiwari2015dipole} find $1110 \pm 370$ [km/s] using the NVSS catalog \citep{condon1998nrao} and \cite{colin2017high} find the much larger $1729 \pm 187$ [km/s] by combining the NVSS and SUMSS catalog \citep{mauch2003sumss}. \cite{bengaly2019testing}  note that linear estimators, used in many but not all radio dipole estimations, may lead to biased estimation of the amplitude and direction.   \cite{siewert2021cosmic}  using a quadratic estimator performed a reanalysis of four radio catalogs (TGSS-ADR1 \citep{intema2017gmrt}, WENSS \citep{de2000vizier}, SUMSS, and NVSS), and found consistent results with previous analyses, a direction in agreement with the CMB dipole but a larger amplitude.

In addition to the CMB and radio source counts there are numerous other manner to constrain the kinematic dipole such as the thermal Sunyaev-Zel'dovich \citep{akrami2020planck}, the X-ray background \citep{plionis1999rosat}, galaxy cluster scaling relations \citep{migkas2021cosmological} and supernovae observations \citep{colin2019evidence}. The latter two methods also find tensions with the CMB kinematic dipole measurement.

It is also interesting to consider how local structures will magnify and shift angular positions of distant radio sources in a non-isotropic manner such that this signal could be confused with the kinematic dipole. The effects of kinematic aberration may be confused with the effects of lensing such that in many analyses (\cite{tiwari2015dipole}, \cite{akrami2020planck}, \cite{aghanim2014planck}, \cite{ferreira2020first}) the local lensing dipole has to be assumed to be small, by constraining the dipole in this work and constraining the dipole to be small, this is no longer an assumption.

In Section \ref{section:radio_dipole} we show how the motion of an observer and lensing changes the observed radio source counts on the sky. In Section \ref{section:local} we discuss constraints on the lensing dipole using the density field reconstructed from low-redshift galaxy surveys. In Section \ref{section:data} we present the data used in our analysis and measurements of the lensing and kinematic dipoles in radio source counts before concluding in Section \ref{section:conclusion}.

\section{ Radio source count dipole }
\label{section:radio_dipole}

The lensing dipole increases the flux of incoming radiation through lensing magnification, therefore introducing faint sources into a magnitude limited sample and deflects the path of photons away from the position of the dipole, therefore reducing the number of sources such that these two effects act against one another. The lensing deflection is at some level degenerate with the aberration due to our motion.

Here we consider the combined effects of kinematic and lensing dipole. Doppler boosting will shift the observed frequency of radiation in comparison to an observer in the rest frame of the source such that the observer frequency, $\nu_{\text{obs}}$, can be related to the rest frame frequency, $\nu_{\text{rest}}$,

\begin{equation}
    \nu_{\text{obs}} \approx \nu_{\text{rest}} ( 1 + \beta \cos \theta )
\end{equation}

to first order in $\beta$, where $\beta=v/c$ for an observer moving at speed $v$ at an angle $\theta$ with respect to the source rest frame.

As the flux density of radio sources follows a power-law in frequency, $f \propto \nu^{-\alpha}$, with $\alpha \approx 0.76$ (\cite{tiwari2019radio}), the frequency shift will change the measured flux density at a given observed frequency. 

The combined flux density change from Doppler boosting and lensing magnification is,

\begin{equation}
f_{\text{obs}} = f_{\text{int}} ( 1 + \beta )^{1+\alpha} ( 1 + 2 \kappa ) 
\end{equation}

where $f_{\text{obs}}$ is the observed source flux density, $f_{\text{int}}$ is the intrinsic source flux density in the absence of lensing and Doppler boosting and $\kappa$ is the lensing convergence. The observed flux is increased by lensing due to the increase of the solid angle of the source, see for example \cite{schmidt2009size}, \cite{broadhurst1994mapping}.  Additionally the solid angle on the sky is changed by aberration from the observer motion and the lensing magnification as,

\begin{equation}
d \Omega_{\text{obs}} = d\Omega_{\text{int}} ( 1 + \beta)^{-2}  ( 1 + 2 \kappa ).
\end{equation}

Note that a positive lensing dipole increases the angle on the sky, whereas a positive kinematic dipole will reduce it. Combining these two effects we can consider the change in number counts at a position $\vec{\theta}$ for a flux limited survey,

\begin{equation}
n_{\text{obs}}(\vec{\theta}) = n_{\text{int}}(\vec{\theta}) \left( 1 + [ 2 + x ( 1 + \alpha )] \beta \cos(\theta) +  2[ x - 1 ] \kappa(\vec{\theta}) \right)
\end{equation}
\label{equation:n}

where $x$ is related to the slope of source flux function at the flux limit,

\begin{equation}
x = \frac{d \ln \text{ n }_{\text{int}} }{ d \ln \, f} |_{f_{\text{lim}}}
\end{equation}
\label{equation:x}

this quantity can be measured on the data. Strictly we should use the intrinsic source population instead of osberved source population after the effects of Doppler boosting and lensing, fortunately the difference between the two will be small provided $x$ varies slowly with flux.

We see that in Equation \ref{equation:n} there is a term proportional to the lensing convergence, $\kappa$, this is the term which is normally neglected in radio source count dipole analyses.

\section{ Local lensing dipole with Cosmic Flows }
\label{section:local}

The anticipated local lensing dipole may be estimated using low-redshift galaxy surveys. We use the local luminosity-weighted galaxy density field, $\delta_g$, provided by \citep{carrick2015cosmological} \footnote{https://cosmicflows.iap.fr/} which extends out to $200$ [Mpc/h] we can attempt to estimate the lensing contribution of such structures. This luminosity-weighted galaxy density field was reconstructed using the 2M++ full sky galaxy redhsift catalog \citep{lavaux20112m}.

In order to convert the galaxy density into a matter density we assume the bias $b=1.2$ which we obtain from \cite{carrick2015cosmological} and taking $\Omega_m =0.3$. Taking a linear biasing relation the galaxy density is related to dark matter density as $\delta_g = b \delta_m$ where the density contrast $\delta_m = \rho_m / \bar{\rho}_m -1$ with $\bar{\rho}_m$ the mean matter density of the universe. Subsequently we reconstruct the lensing convergence, $\kappa$, along the line of sight using the born-approximation and ray-tracing through the provided cosmic-flows density box with the following equation,

\begin{equation}
\kappa( \theta  ) = \frac{3}{2} \frac{H_0^2}{c^2} \Omega_m \int_0^{r_{\text{max}}} dr' \frac{r'(r-r')}{r} \frac{\delta_m(\theta,r')}{a(r')}
\end{equation}

where $r$ is the comoving distance, $H_0$ is the Hubble constant, $\Omega_m$ is the matter density and $a$ is the scale factor. These measurements are very noisy and we are unable to detect any dipolar structure in the resultant projected convergence map. This is also unsurprising in that \cite{carrick2015cosmological} do not find evidence of a local void found elsewhere, unlike more recent cosmic-flows results (\cite{tully2019cosmicflows}). As these are not yet public we have used the older results.

\section{ Joint constraints on kinetic and lensing dipoles from radio source counts }
\label{section:data}

In this section we constrain the kinematic and lensing dipole using radio source count dipole measurements. The two effects are separated by using a series of different flux density cuts.

\subsection{Data}

We are able to place constraints on the lensing dipole and the kinematic dipole using the results of \cite{tiwari2015dipole}. They have measured the dipole anisotropy of radio source counts using the NRAO VLA Sky Survey (NVSS) \citep{condon1998nrao}. The NVSS is a radio continuum survey operating at the frequency $1.4$ GHz and covering the entire northern sky. The median redshift of the radio sources is $z \sim 1$ at which point the effects of local structures should be negligible although they also include additional cuts which eliminates the supergalactic plane. They find consistent results for the dipole amplitude regardless of which cuts are made, indicating the contamination from local structures is negligible.

They find a much larger peculiar velocity than the CMB with $v = 1110 \pm 370$ [km/s]. Using the measured dipole for each flux density cut we break the degeneracy between the lensing and kinematic dipole. These measurement are summarised in Table \ref{table:tiwari}, which with addition of the measurement of $\alpha \approx 0.76$, is all that is needed to estimate both the lensing and kinematic dipole.

\begin{table}
\begin{center}
 \begin{tabular}{||c c c c c||} 
 \hline
\protect$f_{ \rm min}$ [mJy] & \protect$x$ & \protect$\mathcal{D}$  & \protect$\sigma_{\mathcal{D}}$ & \protect$\text{N}_{\text{sources}} $\\ [0.5ex] 
 \hline\hline
 10 & 0.902 & 0.0096 & 0.0026 & $4.0 \cdot 10^5$ \\ 
 \hline
 20 & 1.006 & 0.0125 & 0.0040 & $2.4 \cdot 10^5$ \\
 \hline
 30 & 1.072 & 0.0143 & 0.0048 & $1.6 \cdot 10^5$\\
 \hline
 40 & 1.123 & 0.0136 & 0.0049 & $1.6 \cdot 10^5$ \\
 \hline
 50 & 1.168 & 0.0157 & 0.0059 & $1.0 \cdot 10^5$ \\ [1ex] 
 \hline
\end{tabular}
\caption{These measurements are from \protect\cite{tiwari2015dipole} Table 1. and Table 6. set (b). \protect$f_{\rm min}$ is the flux density cut used in the radio dipole measurement, \protect$x$ is the slope of the source flux function defined in Equation \protect\ref{equation:x}, \protect$\mathcal{D}$ is the dipole amplitude defined in Equation \protect\ref{equation:dipole}, \protect$\sigma_\mathcal{D}$ is the uncertainty on the measured dipole and  \protect$\text{N}_{\text{sources}} $ is the number of radio sources used for the dipole measurement.}
\end{center}
\label{table:tiwari}
\end{table}

\subsection{Model}

In contrast to previous work we estimate the magnitude of the lensing dipole at the same time as the kinematic dipole using the equation,

\begin{equation}
\mathcal{D} = [ 2 + x ( 1 + \alpha )] \beta +  2[ x - 1 ] \kappa
\label{equation:dipole}
\end{equation}

as the value of $x$ changes for different flux density cuts these two quantities may be constrained at the same time. We further make the simplification that the lensing dipole and kinematic dipole are aligned, such an approximation is justified as the same structures inducing the kinematic dipole should dominate the lensing dipole. 

\subsection{Results}

\begin{figure}
\centering
\includegraphics[width=9cm]{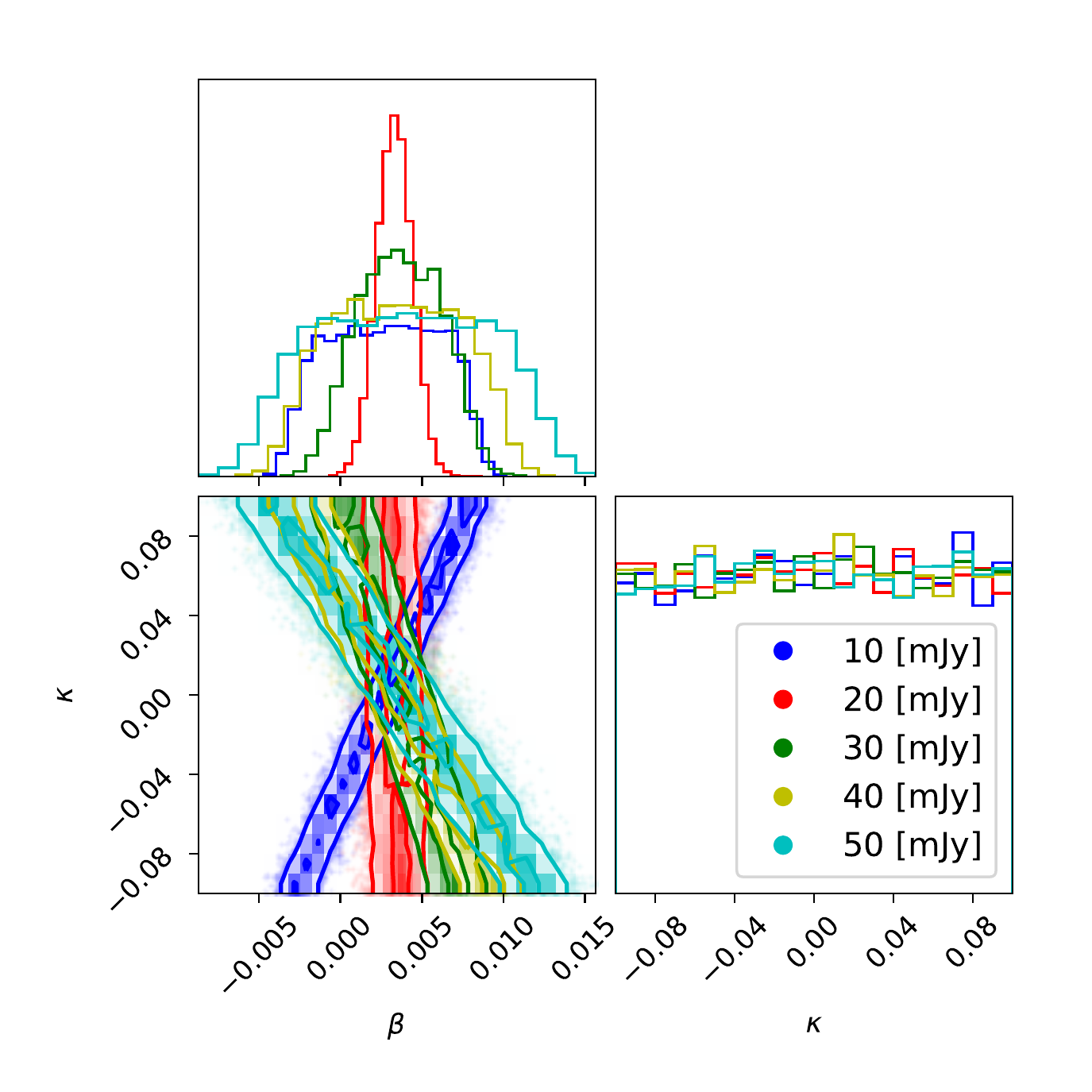}
\caption{ Constraints on the kinematic and lensing dipoles from individual flux density cut measurements, where the contours show the $1\sigma$ and $2\sigma$ levels. Unsurprisingly the constraints are degenerate and the constraints on the kinematic dipole are due to the prior $|\kappa|<10^{-1}$. This plot was made using the corner python package \protect\citep{corner} \protect\footnotemark. }
\label{image:indv_corner_plot}
\end{figure}

Firstly we constrain the lensing and kinematic dipoles separately for each individual flux density cut with the prior that $|\kappa|<10^{-1}$. Using a Gaussian likelihood and an MCMC analysis using the python package emcee \citep{emcee} \footnote{https://emcee.readthedocs.io/}. These constraints are shown in Figure \ref{image:indv_corner_plot}. Unsurprisingly $\kappa$ and $\beta$ are degenerate such that the constraint only arises due to the $\kappa$ prior. We also see that the orientation of the degeneracy changes with the flux density limit such that the combination of these measurement can constrain both of the dipoles. It is worth noting that for the flux density cut at $f_{\text{min}} = 20$ [mJy], the red contour in Figure \ref{image:indv_corner_plot}, the source count dipole is not very sensitive to the lensing dipole, this is because $x \approx 1$ at which point the effects of lensing magnification and the change in solid angle on the sky due to lensing cancel one another.

\footnotetext{https://corner.readthedocs.io/}

Secondly we perform a combined analysis which allows us to break the degeneracy between $\kappa$ and $\beta$, in this case we no longer require a prior on $\kappa$.  The measurements of the dipole at each flux density cut are not independent, as the sources overlap. In order to account for this overlap we use a multivariate Gaussian likelihood and calculate the error for each component of the covariance matrix as, $S_{ij} = r \sigma_i \sigma_j$, where the correlation coefficent $r$ is simply the ratio of the overlap in the number of radio sources used in each measurement. In practice the errors do not significantly change when each measurement is considered to be independent. This should be considered a cautious approach given that the constraining power of each sample happens for sources at the flux density limit, naturally this is information not shared between any of flux density cuts. 

 We take into account errors on $\alpha$ and $x$ by including them as nuisance parameters with Gaussian priors in our MCMC analysis. For $\alpha$ we take its mean to be, as stated earlier $\alpha=0.76$ with the conservative error  $\sigma_\alpha=0.2$  \citep{tiwari2019radio}.  For $x$ we use the values in  Table \ref{table:tiwari} with errors $\sigma_x = x/\sqrt{N_{\rm sources}}$. The inclusion of the errors on $\alpha$ and $x$ leads to less than a 5 \% increase in the errors on $\beta$ and $\kappa$. Additionally  \cite{tiwari2019radio}  find a slight increase of $\alpha$ with flux density, with the large uncertainty used on $\alpha$ in our analysis we will not be sensitive to these effects.

\begin{figure}
\centering
\includegraphics[width=9cm]{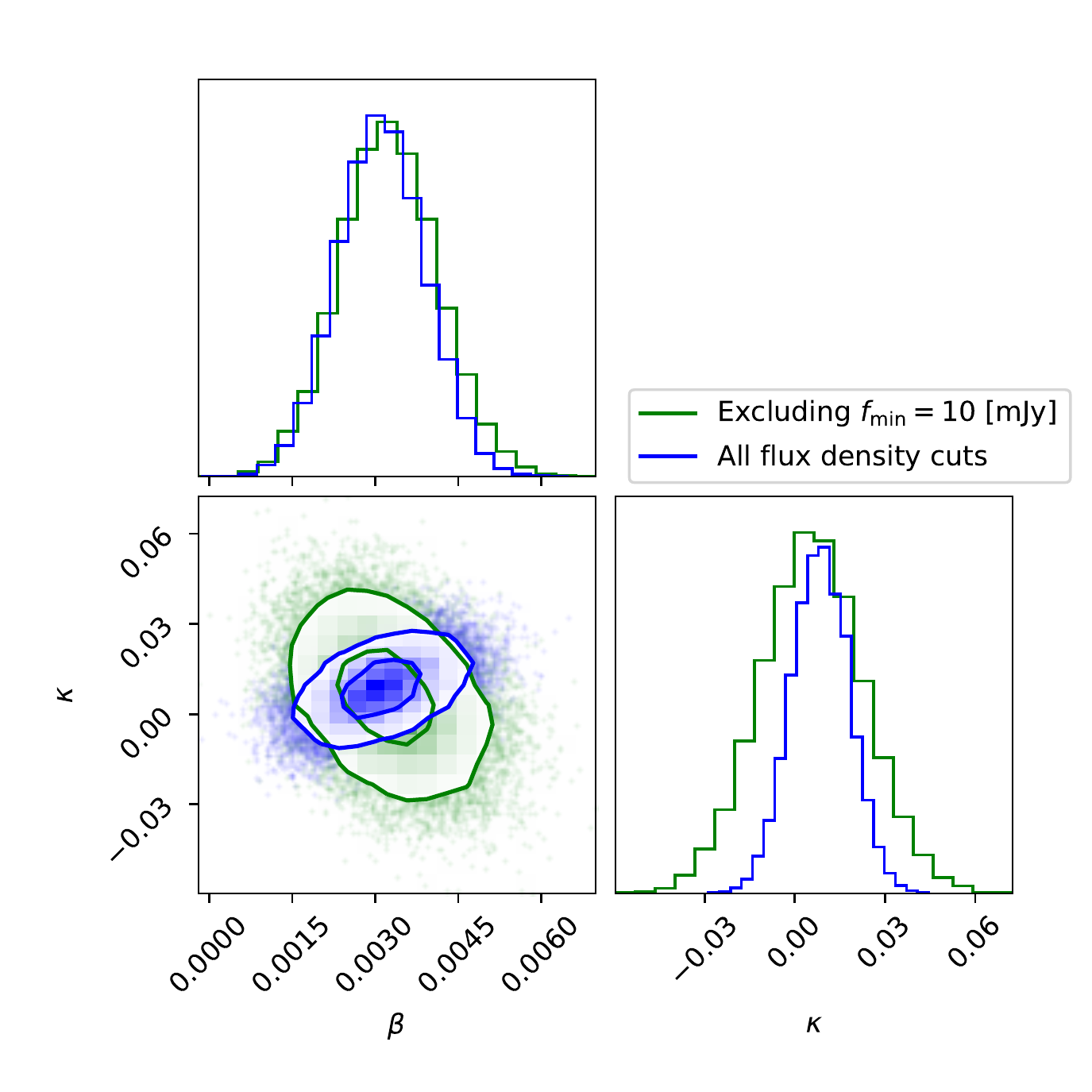}
\caption{ Constraints on the kinematic and lensing dipoles from the combination of dipole measurements at different flux density cuts, where the contours show the $1\sigma$ and $2\sigma$ levels. In blue the analysis using every flux density cut and in green every flux density cut excluding the 10 [mJy] flux density cut which may suffer from systematic effects. We see that there is little degeneracy left between the two parameters. This plot was made using the corner python package \citep{corner}. }
\label{image:corner_plot}
\end{figure}

The results of the MCMC analysis, are shown in Figure \ref{image:corner_plot}. We obtain $v=930 \pm 240$ [km/s] and $\kappa = 9 \cdot 10^{-3} \pm 10 \cdot 10^{-3}$ , a slightly lower speed than the results of \cite{tiwari2015dipole}. We are able to place constraints on the lensing dipole finding that $\kappa < 3 \cdot 10^{-2}$ at the $2 \sigma$ level. The inclusion of a lensing dipole does not help to ease the tension between the radio counts and CMB results.

We also performed an analysis excluding the 10 [mJy] flux density threshold. The motivation is that \citep{blake2002detection} found that the NVSS survey may suffer from systematic issue below 15 [mJy]. Excluding this flux cut notably reduces the constraints on the lensing dipole, this is to be expected as seen in Figure  \ref{image:indv_corner_plot} the posterior for the 10 [mJy] is orientated against the other flux density thresholds. We obtain $v=978 \pm 280$ [km/s] and $\kappa = 6 \cdot 10^{-3} \pm 17 \cdot 10^{-3}$.  

\section{Discussion and Conclusions}
\label{section:conclusion}

We have seen that we can constrain at the same time the lensing dipole and the kinematic dipole without degradation of the kinematic dipole constraints. Future surveys such as SKA \citep{bacon2020cosmology} in the radio and Rubin \citep{ivezic2019lsst} and Euclid \citep{laureijs2011euclid} in the optical will deliver large scale extragalactic catalogs with more than a billion sources at high redshifts. An improvement of four orders of magnitudes over the extragalactic catalogs used in this study. We can make a simple prediction on lensing dipole constrains with such surveys by considering the sky fraction, $f_{\text{sky}}$, covered and the number of sources, $N_{\text{s}}$.

Considering only Poisson noise, the dominant source of noise, the variance on the measurement of the dipole is given by \citep{nadolny2021new},

\begin{equation}
    \sigma^2_{\mathcal{D}} \approx \frac{9 }{ N_{\text{s}} f_{\text{sky}} }.
\end{equation}

For SKA there should be on the order of $N_\text{s} = 10^9$ sources and a sky fraction $f_{\text{sky}} \approx 0.63$ which gives a dipole measurement uncertainty of $\sigma_{\mathcal{D}} \approx 10^{-4}$. Therefore assuming the lensing effects and kinematic can be completely separated, the lensing contribution is $\mathcal{D} \approx 0.1 \kappa$, which would imply $\sigma_{\kappa} \approx 1 \cdot 10^{-3}$. Therefore for SKA $|\kappa| < 2 \cdot 10^{-3}$ at the $2 \sigma$ level. Similar but independent constraints will be possible with Rubin and Euclid. Detailed forecasts for the kinematic and the intrinsic clustering dipole for future surveys are provided in \cite{nadolny2021new}.

To conclude, the number counts of sources on the sky will be affected by a lensing dipole induced by local structures. Such a dipole contribution is not normally considered or assumed to be small. In this work we are able to place constraints on the size of the dipole and indeed we have shown that it is small. 

\section*{Acknowledgements}

Calum Murray would like to thank Céline Combet for reading the manuscript and for suggestions which improved the quality of the article. Calum Murray also acknowledges support from the Labex Enigmass.

\section*{Data Availability}
The data needed to reproduce the main results of this article are found in Table \ref{table:tiwari} which have been obtained by \cite{tiwari2015dipole} using the NVSS catalog. The data used in Section \ref{section:local} can be found at https://cosmicflows.iap.fr/.




\bibliographystyle{mnras}
\bibliography{example} 

\begin{thebibliography}{}
\makeatletter
\relax
\def\mn@urlcharsother{\let\do\@makeother \do\$\do\&\do\#\do\^\do\_\do\%\do\~}
\def\mn@doi{\begingroup\mn@urlcharsother \@ifnextchar [ {\mn@doi@}
  {\mn@doi@[]}}
\def\mn@doi@[#1]#2{\def\@tempa{#1}\ifx\@tempa\@empty \href
  {http://dx.doi.org/#2} {doi:#2}\else \href {http://dx.doi.org/#2} {#1}\fi
  \endgroup}
\def\mn@eprint#1#2{\mn@eprint@#1:#2::\@nil}
\def\mn@eprint@arXiv#1{\href {http://arxiv.org/abs/#1} {{\tt arXiv:#1}}}
\def\mn@eprint@dblp#1{\href {http://dblp.uni-trier.de/rec/bibtex/#1.xml}
  {dblp:#1}}
\def\mn@eprint@#1:#2:#3:#4\@nil{\def\@tempa {#1}\def\@tempb {#2}\def\@tempc
  {#3}\ifx \@tempc \@empty \let \@tempc \@tempb \let \@tempb \@tempa \fi \ifx
  \@tempb \@empty \def\@tempb {arXiv}\fi \@ifundefined
  {mn@eprint@\@tempb}{\@tempb:\@tempc}{\expandafter \expandafter \csname
  mn@eprint@\@tempb\endcsname \expandafter{\@tempc}}}

\bibitem[\protect\citeauthoryear{Bacon et~al.,}{Bacon
  et~al.}{2020}]{bacon2020cosmology}
Bacon D.~J.,  et~al., 2020, Publications of the Astronomical Society of
  Australia, 37

\bibitem[\protect\citeauthoryear{Bengaly, Siewert, Schwarz  \&
  Maartens}{Bengaly et~al.}{2019}]{bengaly2019testing}
Bengaly C.~A.,  Siewert T.~M.,  Schwarz D.~J.,   Maartens R.,  2019, Monthly
  Notices of the Royal Astronomical Society, 486, 1350

\bibitem[\protect\citeauthoryear{Blake \& Wall}{Blake \&
  Wall}{2002}]{blake2002detection}
Blake C.,  Wall J.,  2002, arXiv preprint astro-ph/0203385

\bibitem[\protect\citeauthoryear{Broadhurst, Taylor  \& Peacock}{Broadhurst
  et~al.}{1994}]{broadhurst1994mapping}
Broadhurst T.~J.,  Taylor A.,   Peacock J.,  1994, arXiv preprint
  astro-ph/9406052

\bibitem[\protect\citeauthoryear{Burles \& Rappaport}{Burles \&
  Rappaport}{2006}]{burles2006detecting}
Burles S.,  Rappaport S.,  2006, The Astrophysical Journal Letters, 641, L1

\bibitem[\protect\citeauthoryear{Carrick, Turnbull, Lavaux  \& Hudson}{Carrick
  et~al.}{2015}]{carrick2015cosmological}
Carrick J.,  Turnbull S.~J.,  Lavaux G.,   Hudson M.~J.,  2015, Monthly Notices
  of the Royal Astronomical Society, 450, 317

\bibitem[\protect\citeauthoryear{Challinor \& van Leeuwen}{Challinor \& van
  Leeuwen}{2002}]{challinor2002peculiar}
Challinor A.,  van Leeuwen F.,  2002, Physical Review D, 65, 103001

\bibitem[\protect\citeauthoryear{Colin, Mohayaee, Rameez  \& Sarkar}{Colin
  et~al.}{2017}]{colin2017high}
Colin J.,  Mohayaee R.,  Rameez M.,   Sarkar S.,  2017, Monthly Notices of the
  Royal Astronomical Society, 471, 1045

\bibitem[\protect\citeauthoryear{Colin, Mohayaee, Rameez  \& Sarkar}{Colin
  et~al.}{2019}]{colin2019evidence}
Colin J.,  Mohayaee R.,  Rameez M.,   Sarkar S.,  2019, Astronomy \&
  Astrophysics, 631, L13

\bibitem[\protect\citeauthoryear{Condon, Cotton, Greisen, Yin, Perley, Taylor
  \& Broderick}{Condon et~al.}{1998}]{condon1998nrao}
Condon J.~J.,  Cotton W.,  Greisen E.,  Yin Q.,  Perley R.~A.,  Taylor G.,
  Broderick J.,  1998, The Astronomical Journal, 115, 1693

\bibitem[\protect\citeauthoryear{Crawford}{Crawford}{2009}]{crawford2009detecting}
Crawford F.,  2009, The Astrophysical Journal, 692, 887

\bibitem[\protect\citeauthoryear{Ellis \& Baldwin}{Ellis \&
  Baldwin}{1984}]{ellis1984expected}
Ellis G.,  Baldwin J.,  1984, Monthly Notices of the Royal Astronomical
  Society, 206, 377

\bibitem[\protect\citeauthoryear{Ferreira \& Quartin}{Ferreira \&
  Quartin}{2020}]{ferreira2020first}
Ferreira P. d.~S.,  Quartin M.,  2020, arXiv preprint arXiv:2011.08385

\bibitem[\protect\citeauthoryear{Foreman-Mackey}{Foreman-Mackey}{2016}]{corner}
Foreman-Mackey D.,  2016, \mn@doi [The Journal of Open Source Software]
  {10.21105/joss.00024}, 1, 24

\bibitem[\protect\citeauthoryear{{Foreman-Mackey}, {Hogg}, {Lang}  \&
  {Goodman}}{{Foreman-Mackey} et~al.}{2013}]{emcee}
{Foreman-Mackey} D.,  {Hogg} D.~W.,  {Lang} D.,   {Goodman} J.,  2013, \mn@doi
  [\pasp] {10.1086/670067}, \href
  {https://ui.adsabs.harvard.edu/abs/2013PASP..125..306F} {125, 306}

\bibitem[\protect\citeauthoryear{Gibelyou \& Huterer}{Gibelyou \&
  Huterer}{2012}]{gibelyou2012dipoles}
Gibelyou C.,  Huterer D.,  2012, Monthly Notices of the Royal Astronomical
  Society, 427, 1994

\bibitem[\protect\citeauthoryear{I}{I}{2020}]{collaboration2020planck}
I P.~C.,  2020, Astronomy \& Astrophysics, 641, A1

\bibitem[\protect\citeauthoryear{Intema, Jagannathan, Mooley  \& Frail}{Intema
  et~al.}{2017}]{intema2017gmrt}
Intema H.,  Jagannathan P.,  Mooley K.,   Frail D.,  2017, Astronomy \&
  Astrophysics, 598, A78

\bibitem[\protect\citeauthoryear{Ivezi{\'c} et~al.,}{Ivezi{\'c}
  et~al.}{2019}]{ivezic2019lsst}
Ivezi{\'c} {\v{Z}}.,  et~al., 2019, The Astrophysical Journal, 873, 111

\bibitem[\protect\citeauthoryear{Langlois \& Piran}{Langlois \&
  Piran}{1995}]{langlois1995dipole}
Langlois D.,  Piran T.,  1995, arXiv preprint astro-ph/9507094

\bibitem[\protect\citeauthoryear{Laureijs et~al.,}{Laureijs
  et~al.}{2011}]{laureijs2011euclid}
Laureijs R.,  et~al., 2011, arXiv preprint arXiv:1110.3193

\bibitem[\protect\citeauthoryear{Lavaux \& Hudson}{Lavaux \&
  Hudson}{2011}]{lavaux20112m}
Lavaux G.,  Hudson M.~J.,  2011, Monthly Notices of the Royal Astronomical
  Society, 416, 2840

\bibitem[\protect\citeauthoryear{Mauch, Murphy, Buttery, Curran, Hunstead,
  Piestrzynski, Robertson  \& Sadler}{Mauch et~al.}{2003}]{mauch2003sumss}
Mauch T.,  Murphy T.,  Buttery H.,  Curran J.,  Hunstead R.,  Piestrzynski B.,
  Robertson J.,   Sadler E.,  2003, Monthly Notices of the Royal Astronomical
  Society, 342, 1117

\bibitem[\protect\citeauthoryear{Migkas, Pacaud, Schellenberger, Erler,
  Nguyen-Dang, Reiprich, Ramos-Ceja  \& Lovisari}{Migkas
  et~al.}{2021}]{migkas2021cosmological}
Migkas K.,  Pacaud F.,  Schellenberger G.,  Erler J.,  Nguyen-Dang N.,
  Reiprich T.,  Ramos-Ceja M.,   Lovisari L.,  2021, Astronomy \& Astrophysics,
  649, A151

\bibitem[\protect\citeauthoryear{Nadolny, Durrer, Kunz  \& Padmanabhan}{Nadolny
  et~al.}{2021}]{nadolny2021new}
Nadolny T.,  Durrer R.,  Kunz M.,   Padmanabhan H.,  2021, arXiv preprint
  arXiv:2106.05284

\bibitem[\protect\citeauthoryear{{Planck Collaboration LVI}}{{Planck
  Collaboration LVI}}{2020}]{akrami2020planck}
{Planck Collaboration LVI} 2020, Astronomy and Astrophysics (A \& A), 644

\bibitem[\protect\citeauthoryear{{Planck Collaboration XXVII}}{{Planck
  Collaboration XXVII}}{2014}]{aghanim2014planck}
{Planck Collaboration XXVII} 2014, Astronomy \& Astrophysics, 571, A27

\bibitem[\protect\citeauthoryear{Plionis \& Georgantopoulos}{Plionis \&
  Georgantopoulos}{1999}]{plionis1999rosat}
Plionis M.,  Georgantopoulos I.,  1999, Monthly Notices of the Royal
  Astronomical Society, 306, 112

\bibitem[\protect\citeauthoryear{Rubart \& Schwarz}{Rubart \&
  Schwarz}{2013}]{rubart2013cosmic}
Rubart M.,  Schwarz D.~J.,  2013, Astronomy \& Astrophysics, 555, A117

\bibitem[\protect\citeauthoryear{Schmidt, Rozo, Dodelson, Hui  \&
  Sheldon}{Schmidt et~al.}{2009}]{schmidt2009size}
Schmidt F.,  Rozo E.,  Dodelson S.,  Hui L.,   Sheldon E.,  2009, Physical
  Review Letters, 103, 051301

\bibitem[\protect\citeauthoryear{Siewert, Schmidt-Rubart  \& Schwarz}{Siewert
  et~al.}{2021}]{siewert2021cosmic}
Siewert T.~M.,  Schmidt-Rubart M.,   Schwarz D.~J.,  2021, Astronomy \&
  Astrophysics, 653, A9

\bibitem[\protect\citeauthoryear{Singal}{Singal}{2011}]{singal2011large}
Singal A.~K.,  2011, The Astrophysical Journal letters, 742, L23

\bibitem[\protect\citeauthoryear{Tiwari}{Tiwari}{2019}]{tiwari2019radio}
Tiwari P.,  2019, Research in Astronomy and Astrophysics, 19, 096

\bibitem[\protect\citeauthoryear{Tiwari, Kothari, Naskar, Nadkarni-Ghosh  \&
  Jain}{Tiwari et~al.}{2015}]{tiwari2015dipole}
Tiwari P.,  Kothari R.,  Naskar A.,  Nadkarni-Ghosh S.,   Jain P.,  2015,
  Astroparticle Physics, 61, 1

\bibitem[\protect\citeauthoryear{Tully, Pomar{\`e}de, Graziani, Courtois,
  Hoffman  \& Shaya}{Tully et~al.}{2019}]{tully2019cosmicflows}
Tully R.~B.,  Pomar{\`e}de D.,  Graziani R.,  Courtois H.~M.,  Hoffman Y.,
  Shaya E.~J.,  2019, The Astrophysical Journal, 880, 24

\bibitem[\protect\citeauthoryear{de Bruyn et~al.,}{de~Bruyn
  et~al.}{2000}]{de2000vizier}
de Bruyn G.,  et~al., 2000, VizieR Online Data Catalog, pp VIII--62

\makeatother
\end{thebibliography}



\appendix


\bsp	
\label{lastpage}
\end{document}